\documentclass[useAMS,usenatbib]{mn2e}
\pdfoutput=1

\usepackage{graphicx}

\def\EBV{\mbox{$E(B-V)$}}
\def\eBV{\mbox{$e(B-V)$}}
\def\dEBV{\mbox{$\delta\EBV$}}
\def\mEBV{\mbox{$\overline{\dEBV}$}}
\def\mEBV{\mbox{$\langle{\dEBV}\rangle$}}
\def\xEBV{\mbox{$\dEBV_{max}$}}

\def\ss{\mbox{$\mathcal{S}$}}
\def\x2{\mbox{$R_{rms}$}}

\title[Differential reddening in globular clusters]{Mapping the differential reddening 
in globular clusters\thanks{Based on observations with the NASA/ESA Hubble Space Telescope, 
obtained at the Space Telescope Science Institute, which is operated by AURA, Inc., under 
NASA contract NAS5-26555, under programs GO-10775 (PI: A. Sarajedini) and GO\,11586 (PI: 
A. Dotter).}}

\author[C. Bonatto, Fab\'\i ola Campos and S.O. Kepler]{C. Bonatto$^1$, Fab\'\i ola Campos$^1$ 
and S.O. Kepler$^{1}$\\
$^1$ Departamento de Astronomia, Universidade Federal do Rio Grande do Sul, Av. Bento 
Gon\c{c}alves 9500\\ 
Porto Alegre 91501-970, RS, Brazil}

\begin{document}

\pagerange{\pageref{firstpage}--\pageref{lastpage}}

\maketitle

\label{firstpage}

\begin{abstract}
We build differential-reddening maps for 66 Galactic globular clusters (GCs) with 
archival HST WFC/ACS F606W and F814W photometry. Because of the different GC sizes 
(characterised by the half-light radius $R_h$) and distances to the Sun, the WFC/ACS 
field of view ($200\arcsec\times200\arcsec$) coverage ($R_{obs}$) lies in the range 
$1\la R_{obs}/R_h\la15$ for about $85\%$ of the sample, with about 10\% covering only
the inner ($R_{obs}\la R_h$) parts. We divide the WFC/ACS field of view across each 
cluster in a regular cell grid, and extract the stellar-density Hess diagram from each 
cell, shifting it in colour and magnitude along the reddening vector until matching the 
mean diagram. Thus, the maps correspond to the internal dispersion of the reddening 
around the mean. Depending on the number of available stars (i.e. probable members 
with adequate photometric errors), the angular resolution of the maps range from 
$\approx7\arcsec\times7\arcsec$ to $\approx20\arcsec\times20\arcsec$. We detect 
spatially-variable extinction in the 66 globular clusters studied, with mean values 
ranging from $\mEBV\approx0.018$ (NGC\,6981) up to $\mEBV\approx0.16$ (Palomar\,2). 
Differential-reddening correction decreases the observed foreground reddening and the 
apparent distance modulus but, since they are related to the same value of \EBV, the 
distance to the Sun is conserved. Fits to the mean-ridge lines of the highly-extincted 
and photometrically scattered globular cluster Palomar\,2 show that age and metallicity 
also remain unchanged after the differential-reddening correction, but measurement 
uncertainties decrease because of the reduced scatter. The lack of systematic 
variations of \mEBV\ with both the foreground reddening and the sampled cluster area 
indicates that the main source of differential reddening is interstellar. 
\end{abstract}


\section{Introduction}
\label{intro}

Spatially-variable extinction, or differential reddening (hereafter DR), occurs in all 
directions throughout the Galaxy and is also present across the field of view of most
globular clusters (GCs). By introducing non-systematic confusion in the actual colour
and magnitude of member stars located in different parts of a GC, it leads to a broadening 
of evolutionary sequences in the colour-magnitude diagram (CMD). Thus, together with 
significant photometric scatter, unaccounted-for DR hampers the detection of multiple-population 
sequences and may difficult the precise determination of fundamental parameters, especially 
the age, metallicity and distance of GCs. 

Over the years, several approaches have been used to map DR across the field of particular 
GCs, mostly focusing on rather restricted stellar evolutionary stages of the CMD. A rather 
comprehensive review of early works can be found in Alonso-Garc\'\i a et al. (2011, 2012). 
More recently, \citet{Milone2012a} study the DR of 59 GCs observed with the Hubble Space 
Telescope (HST) WFC/ACS. Their method involves defining a fiducial line for the main-sequence 
(MS) and measuring the displacement (along the reddening vector) in apparent distance modulus 
and colour for all MS stars. This correction proved to be important 
for a better characterization of multiple sub-giant branch populations in NGC\,362, NGC\,5286, 
NGC\,6656, NGC\,6715, and NGC\,7089 (\citealt{Piotto2012}). Alonso-Garc\'\i a et al. (2011, 2012) 
search for DR in several GCs observed with the Magellan 6.5m telescope and HST. Their approach is 
similar to that of \citet{Milone2012a} but, instead of restricting to the MS, they build a mean 
ridge line (MRL) for most of the evolutionary stages expected of GC stars. Along similar lines, 
\citet{Tz5} use HST WFC/ACS photometry to estimate the DR in the highly-extincted stellar system 
Terzan\,5, which presents an average (mostly foreground) colour excess of $\EBV=2.38$. They build 
a regular grid of $25\times25$ cells across the field of view of Terzan\,5 and search for spatial 
variations, with respect to the bluest cell, in the mean colour and magnitude of MS stars extracted 
within each cell. The cell to cell differences in the mean (apparent distance modulus and colour) 
values are entirely attributed to internal reddening variations, which they find to reach a dispersion 
of $\EBV\sim0.67$ around the mean.

In the present paper we use archival HST WFC/ACS data to estimate the differential reddening in 
a large sample of Galactic GCs. Our approach incorporates some of the features of the previous 
works, while including others that we propose to represent improvements. For instance, we use 
all the evolutionary sequences present in the CMD of a GC, take the photometric uncertainties 
explicitly into account, and exclude obvious non-member stars to build the Hess diagram. Thus, 
instead of defining a fiducial line, we use the actual stellar density distribution present in 
the Hess diagram of all stars as the template against which internal reddening variations are 
searched for. In short, we divide the on-the-sky projected stellar distribution in a spatial 
grid, extract the CMD of each cell, convert it into the Hess diagram, and find the reddening 
value that leads to the best match of this cell with the template Hess diagram. The final 
product is a DR map with a projected spatial resolution directly related to the number of 
stars available in the image. This, in turn, can be used to obtain the DR-corrected CMD.

This paper is organised as follows: in Sect.~\ref{sample} we briefly discuss the data and 
the photometric cuts applied. In Sect.~\ref{approach} we describe the approach adopted. 
In Sect.~\ref{results} we discuss the results for the GC sample. Concluding remarks are given 
in Sect.~\ref{Conclu}.

\section{The GC sample}
\label{sample}

Most of the GCs focused here are part of the HST WFC/ACS sample obtained under program number 
GO\,10775, with A. Sarajedini as PI, which guarantees some uniformity on data reduction and 
photometric quality. GO\,10775 is a HST Treasury project in which 66 GCs were observed through 
the F606W and F814W filters. For symmetry reasons, we work only with GCs centred on the images.
Also, for statistical reasons (Sect.~\ref{approach}), our approach requires a minimum of 
$\sim3000$ stars (probable members with adequate colour uncertainty) available in the CMD. 
Both restrictions reduced the sample to 60 GCs. In addition, we include the 6 GCs (Pyxis, 
Ruprecht\,106, IC\,4499, NGC\,6426, NGC\,7006, and Palomar\,15) observed with WFC/ACS under 
program GO\,11586, with A. Dotter as PI. These 6 GCs satisfy our symmetry/number of stars 
conditions. Further details on the observations and data reduction are in \citet{Milone2012a} 
and \citet{6Dotter}, respectively for both samples. The final sample contains 66 GCs.

The field of view of WFC/ACS is $\approx200\arcsec\times200\arcsec$, which implies that, 
depending on the GC distance and intrinsic size, different fractions of the angular distribution
of cluster stars 
are covered by the observations. We characterize this effect by measuring the ratio between the 
maximum separation reached by WFC/ACS ($\approx100\arcsec$) computed at the GC distance and its 
half-light radius (taken from the 2010 update of \citealt{H96} - hereafter H10), $\ss\equiv 
R_{obs}/R_h$. The results are given in Table~\ref{tab1}. A first-order analysis of the structural 
parameters in H10 shows that the tidal and half-light radii relate linearly as $R_t\sim10\,R_h$. 
Thus, in  most cases we are sampling the very inner GC parts, since $\approx10\%$ of our sample 
has $\ss\le1$, such as in the case of NGC\,6397 and NGC\,6656; most of the sample, $\approx85\%$, 
lies in the range spanned by $1\le\ss\le15$, with 4 GCs (NGC\,4147, Palomar\,15, Palomar\,2, 
NGC\,7006) reaching higher coverage levels.

Still regarding the photometric quality, \citet{Anders08} show that unmodelable PSF variations 
introduce a slight shift in the photometric zero point as a function of the star’s location in the 
field. The shifts cancel out when large areas are considered, but may be as large as $\pm0.02$ mag 
in colour when comparing stars extracted from different, small areas across the field of GCs. In 
addition, they note that this residual variation is very hard to distinguish from actual spatially-variable 
reddening. We'll return to this point in Sect.~\ref{results}.

\begin{figure}
\resizebox{\hsize}{!}{\includegraphics{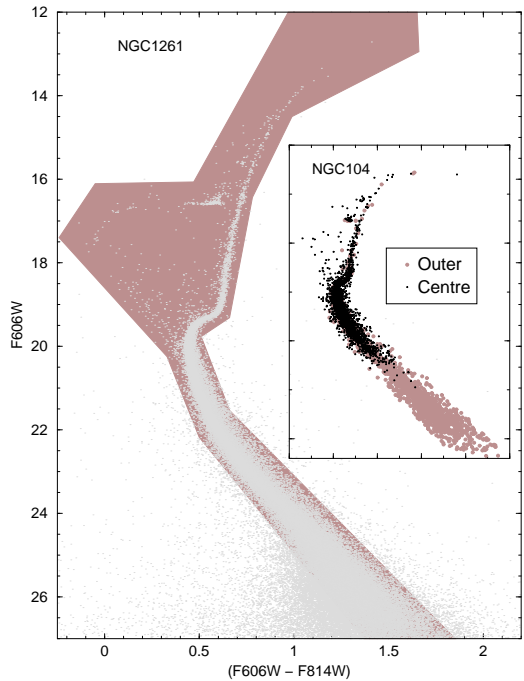}}
\caption{Example of a colour-magnitude filter (shaded polygon) used to minimize field 
contamination on NGC\,1261. The inset shows morphological differences in CMDs extracted 
from the relatively crowded centre and an outer region of NGC\,104.}
\label{fig1}
\end{figure}

Galactic GCs suffer from varying fractions of field-star contamination (\citealt{Milone2012a}).
Since comparison fields obtained with the same instrumentation are not available, we minimise 
non-member contamination in our analysis by applying colour-magnitude filters to exclude obvious 
field stars. These filters are designed to exclude stars with colour and/or magnitude clearly 
discordant from those expected of GC stellar evolutionary sequences. For high-quality photometry, 
these filters are more significant for fainter stars. We designed filters for each GC 
individually, respecting particularities associated with the CMD morphology (e.g. the presence
of a blue horizontal branch or a red clump, blue stragglers, etc.) and photometric 
uncertainties. A typical example of such a filter is shown in Fig.~\ref{fig1} for NGC\,1261.

\begin{figure}
\resizebox{\hsize}{!}{\includegraphics{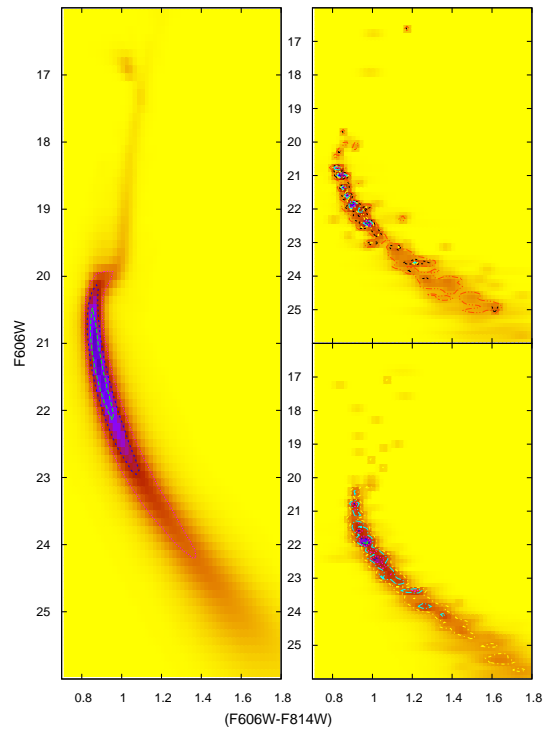}}
\caption{The bluest (top-right panel) and reddest (bottom-right) Hess diagrams of NGC\,6388 
are compared to the mean (observed) one (left). The difference in reddening between both 
amounts to $\dEBV=0.133$.}
\label{fig2}
\end{figure}

Despite the high quality of the sample, photometric uncertainties, which increase towards 
fainter magnitudes, may lead to a significant broadening of the lower main sequence. Thus, 
as an additional photometric quality control, we discard any star with colour uncertainty 
($\sigma_{col}$) higher than a given threshold ($\sigma_{max}$). For statistical reasons, 
we work with $\sigma_{col}$ between 0.2 and 0.4 mag, lower for GCs with a high number of 
available stars ($\ga2\times10^5$) and increasing otherwise. Depending on the photometric 
quality and the amount of field-star contamination, the colour-magnitude filter combined to 
the $\sigma_{col}$ criterion exclude, on average, $\sim15-40\%$ of the stars originally 
available in the CMDs. Parameters of the GC sample are given in Table~\ref{tab1}, where the
Galactic coordinates and the foreground reddening are taken from H10.

\section{The adopted approach}
\label{approach}

The underlying concept is that, if DR is non negligible, CMDs extracted in different parts 
of a GC would be similar to each other but shifted along the reddening vector by an amount 
proportional to the relative extinction between the regions. In this sense, minimization of 
the differences by matching the CMDs can, in principle, be used to search for the extinction 
values affecting different places of a GC.

Obviously, differences between CMDs extracted from various parts of the same GC are expected 
to occur either by dynamical evolution (e.g. mass/luminosity segregation), by the presence of 
multiple populations (e.g. \citealt{Milone2012b} and references therein), or by observational 
limitations (e.g. crowding tends to produce a deficiency of faint stars in the central parts 
of a GC with respect to the outer parts). For instance, the latter effect is clearly illustrated 
by the CMDs of NGC\,104 extracted at the relatively crowded centre and in a ring $\sim100\arcsec$ 
away from it (inset of Fig.~\ref{fig1}), a sparser region that allows detection of fainter stars. 
Then, a direct comparison between a central and an outer CMD might come up with differences that 
could be erroneously interpreted as DR. Thus, to minimize effects unrelated to DR, CMDs extracted 
in different parts of a GC are compared to the mean CMD. Specifically, the mean CMD contains all 
the WFC/ACS image stars that remain after we apply our photometric criteria (Sect.~\ref{sample}). 
Statistically, the mean CMD displays the spatially-averaged colour and magnitude properties of a 
GC, weighted according to the number of stars in each region. In this sense, variations detected 
among CMDs extracted in different parts of a GC should reflect the dispersion of reddening around 
the mean value. In addition, since the mean CMD contains the highest (with respect to the partial 
CMDs) number of stars, this procedure maximizes the statistical significance of the CMD comparison.

\begin{figure}
\resizebox{\hsize}{!}{\includegraphics{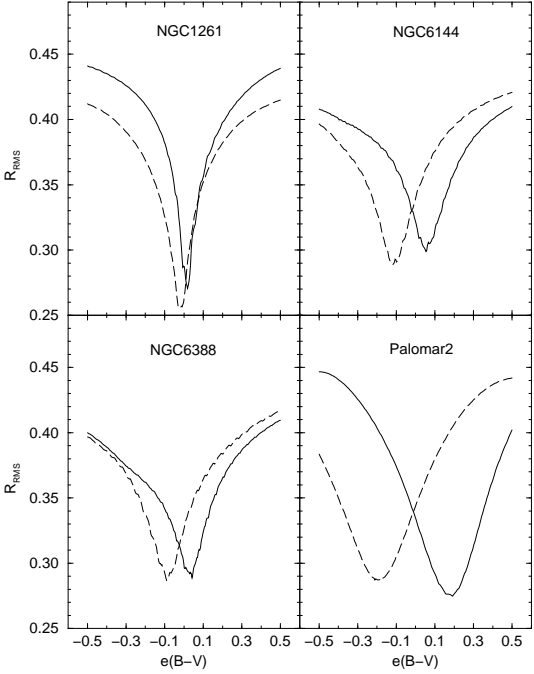}}
\caption{Convergence pattern (\x2\ as a function of \eBV) computed for the bluest (solid line) 
and reddest (dashed) cells of selected GCs.}
\label{fig3}
\end{figure}

\begin{figure}
\resizebox{\hsize}{!}{\includegraphics{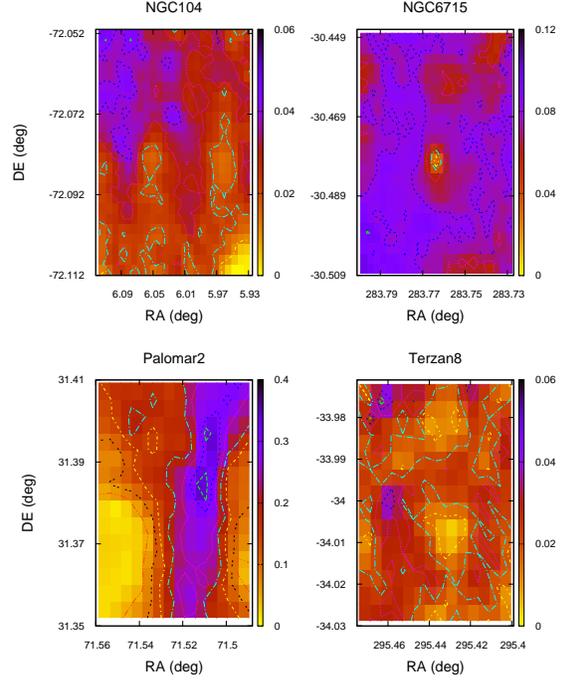}}
\caption{Differential-reddening maps of NGC\,104, NGC\,6715, Palomar\,2 and
Terzan\,8. The range of values of \dEBV\ is shown by the vertical bars.}
\label{fig4}
\end{figure}

However, the direct comparison between CMDs may also introduce some additional problems. 
Among these, we call attention to the discreteness associated with CMDs (especially those 
built with 
a small number of stars) and the remaining photometric uncertainties (see Sect.~\ref{sample}). 
To minimize these effects, we work with the (rather continuous) Hess diagram\footnote{Hess 
diagrams show the relative density of occurrence of stars in different colour-magnitude 
cells of the CMD (\citealt{Hess24}).}. 

Photometric uncertainties - which are assumed to be Normally distributed -  are taken into 
account when building the Hess diagrams. Formally, if the magnitude (or colour) of a given 
star is given by $\bar x\pm\sigma$, the probability of finding it at a specific value $x$ 
is given by $P(x)=\frac{1}{\sqrt{2\pi}\sigma}\,e^{{-\frac{1}{2}}\left(\frac{x-\bar x}
{\sigma}\right)^2}$. Thus, for each star we compute the fraction of the magnitude and 
colour that occurs in a given bin of a Hess diagram, which corresponds to the difference 
of the error functions computed at the bin borders; by definition, the sum of the density 
over all Hess bins is the number of input stars. Since we are comparing cells with unequal 
numbers of stars, the corresponding Hess diagrams are normalized to unity, so that differences
in number counts do not introduce biases into the analysis.

Then, the on-the-sky stellar distribution of the remaining stars ($N_{CMD}$) is divided in 
a grid of cells with a dimension that depends on the number of stars, and the CMD of each 
cell is extracted and converted into a Hess diagram. As an additional quality constraint, 
we only compute DR for cells having a number of stars higher than a minimum value $N_{min}$. 
After some tests, we settled for the range $50\le N_{min}\le150$, higher for the GCs with 
more stars available. The spatial grid dimensions range from $30\times30$ to $10\times10$,
with resolution increasing with the number of stars. Consequently, the angular resolution of 
the DR maps range from $\approx7\arcsec\times7\arcsec$ to $\approx20\arcsec\times20\arcsec$. 
Grid resolution, $N_{min}$, $\sigma_{max}$, and $N_{CMD}$, are listed in Table~\ref{tab1} for 
all GCs. Finally, we search for the optimum value of \dEBV\ that leads to the best match between 
the Hess diagram of a cell and that of the whole distribution. The reddening vector has axes 
with amplitudes corresponding to the absorption relations $A_{F606W}=2.86\dEBV$ and 
$A_{F814W}=1.88\dEBV$, derived according to \citet{CCM89} for a G2V star.

Part of this process is illustrated in Fig.~\ref{fig2}, in which the Hess diagrams of the 
bluest and reddest cells of NGC\,6388 are compared to the overall/mean one. Except for the 
evolved sequences (and the relative stellar density), the bluest and reddest cell diagrams 
have similar morphologies, differing somewhat in magnitude and colour.

As discussed above, our approach assumes that DR plays the dominant role in producing 
variations such as those in Fig.~\ref{fig3} among CMDs extracted in different regions of 
the same GC. Consequently, this reduces the problem to searching for a single parameter, 
namely the internal colour excess \dEBV\ difference between regions. We achieve this by 
minimising the root mean squared residuals (\x2) between the mean ($\overline H$) Hess 
diagram and that extracted from a given cell ($H_{c}$), each composed of $n$ and $m$ colour 
and magnitude bins, where

\begin{equation}
\label{eq1}
\x2=\sqrt{\frac{1}{N_{c}}\sum_{i,j=1}^{n,m}\frac{\left[\overline H(i,j)-H_{c}(i,j)\right]^2}
{\left[\overline H(i,j)+H_{c}(i,j)\right]}}~~~.
\end{equation}

The sum is restricted to non-empty Hess bins, and the normalisation by the total density 
in each bin gives a higher weight to the more populated bins\footnote{In Poisson statistics, 
where the uncertainty of a measurement $\mathcal{N}$ is $\sigma=\sqrt{\mathcal{N}}$, our 
definition of \x2\ is equivalent to the usual $\chi^2$.}. Finally, the squared sum is divided 
by the number of stars in the cell, which makes \x2\ dimensionless, and preserves the number 
statistics when comparing the convergence pattern for cells with unequal numbers of stars. All 
things considered, the relative weight that each evolutionary sequence has in the analysis is 
proportional to the respective Hess stellar density, roughly the number of stars it contains.

To minimise \x2\ we use the global optimisation method known as adaptive simulated 
annealing (ASA), because it is relatively time efficient, robust and capable of finding the 
optimum minimum (e.g. \citealt{ASA}; \citealt{SimASA}; \citealt{GCASA}). Since the CMD cells 
may be somewhat bluer or redder than the mean one, the search starts with the parameter \eBV\ 
being free to vary between $-0.5$ and $+0.5$. The initial trial point is randomly selected within 
this range and the starting value of \x2\ is computed. Then ASA takes a step by changing the 
initial parameter, and a new value of \x2\ is evaluated. Specifically, this implies that all 
stars in the cell have their colour and magnitude changed according to the new value of \eBV. 
By definition, any \x2-decreasing (downhill) move is accepted, with the process repeating 
from this new point. Uphill moves may also be taken, with the decision made by the Metropolis 
(\citealt{Metropolis53}) criterion, which enables ASA to escape from local minima. Variation 
steps become smaller as the minimisation is successful and ASA approaches the global minimum. 

We illustrate the convergence pattern (\x2\ as a function of \eBV) for the bluest and reddest
cells of some representative GCs in Fig.~\ref{fig3}. Regardless of the number of available GC 
stars, low for NGC\,6144 ($\sim2\times10^4$) and higher for Palomar\,2 ($\sim4.2\times10^4$), 
NGC\,1261 ($\sim8.4\times10^4$), and NGC\,6388 ($\sim2.7\times10^5$), the \x2\ shape is similar, 
presenting a conspicuous and rather deep minimum, which is easily retrieved by our approach.

\begin{table*}
\caption[]{General properties and derived parameters of the GC sample}
\label{tab1}
\renewcommand{\tabcolsep}{2.1mm}
\begin{tabular}{lrrccrccccc}
\hline\hline
GC & $\ell$ & $b$ & $\ss$ & $n_{RA}\times n_{DE}$ & $N_{min}$& $\sigma_{max}$ & $N_{CMD}$ & \EBV & \mEBV & \xEBV \\
   &($\degr$)&($\degr$)&&(cells)&(stars)&(mag)&($10^4$ stars)&(mag) & (mag) & (mag)\\
(1) & (2) & (3) & (4) & (5) & (6) & (7) & (8) & (9) & (10) & (11)\\
\hline
NGC\,6441  & $353.53$ & $-5.01$  & 10  & $30\times30$ & 150 & 0.2 & 28.6 & 0.47 & $0.069\pm0.036$ & 0.175 \\
NGC\,6388  & $345.56$ & $-6.74$  & 9.0 & $25\times25$ & 150 & 0.2 & 27.0 & 0.37 & $0.047\pm0.019$ & 0.133 \\
NGC\,5139  & $309.10$ & $14.97$  & 0.5 & $30\times30$ & 150 & 0.2 & 26.9 & 0.12 & $0.028\pm0.009$ & 0.050 \\
NGC\,2808  & $282.19$ & $-11.25$ & 6.0 & $30\times30$ & 150 & 0.2 & 25.8 & 0.22 & $0.028\pm0.011$ & 0.068 \\
NGC\,6715  & $5.61$   & $-14.09$ & 15  & $25\times25$ & 150 & 0.2 & 23.4 & 0.15 & $0.077\pm0.010$ & 0.101 \\
NGC\,7089  & $53.37$  & $-35.77$ & 5.4 & $25\times25$ & 150 & 0.2 & 20.9 & 0.06 & $0.032\pm0.009$ & 0.055 \\
NGC\,7078  & $65.01$  & $-27.31$ & 5.2 & $25\times25$ & 150 & 0.2 & 19.9 & 0.10 & $0.032\pm0.014$ & 0.065 \\
NGC\,5024  & $332.96$ & $79.76$  & 6.8 & $25\times25$ & 150 & 0.2 & 19.9 & 0.02 & $0.030\pm0.009$ & 0.068 \\
NGC\,5286  & $311.61$ & $10.57$  & 8.0 & $25\times25$ & 125 & 0.3 & 17.6 & 0.24 & $0.061\pm0.017$ & 0.112 \\
NGC\,5272  & $42.22$  & $78.71$  & 2.2 & $25\times25$ & 150 & 0.3 & 15.0 & 0.01 & $0.031\pm0.009$ & 0.063 \\
NGC\,104   & $305.89$ & $-44.89$ & 0.7 & $25\times25$ & 125 & 0.3 & 14.8 & 0.04 & $0.028\pm0.009$ & 0.051 \\
NGC\,6205  & $59.01$  & $40.91$  & 2.1 & $25\times25$ & 125 & 0.3 & 13.9 & 0.02 & $0.026\pm0.009$ & 0.054 \\
NGC\,5986  & $337.02$ & $13.27$  & 5.3 & $25\times25$ & 150 & 0.3 & 13.6 & 0.28 & $0.048\pm0.022$ & 0.111 \\
NGC\,6341  & $68.34$  & $34.86$  & 4.1 & $25\times25$ & 125 & 0.3 & 12.3 & 0.02 & $0.030\pm0.010$ & 0.064 \\
NGC\,1851  & $244.51$ & $-35.03$ & 11  & $25\times25$ & 125 & 0.3 & 12.0 & 0.02 & $0.025\pm0.010$ & 0.054 \\
NGC\,6093  & $352.67$ & $19.46$  & 8.2 & $23\times23$ & 100 & 0.3 & 11.3 & 0.18 & $0.039\pm0.015$ & 0.078 \\
NGC\,362   & $301.53$ & $-46.25$ & 5.2 & $23\times23$ & 100 & 0.3 & 10.5 & 0.05 & $0.032\pm0.009$ & 0.056 \\
NGC\,5904  & $3.86$   & $46.80$  & 2.1 & $23\times23$ & 100 & 0.3 & 10.3 & 0.03 & $0.033\pm0.009$ & 0.068 \\
NGC\,6541  & $349.29$ & $-11.19$ & 3.5 & $23\times23$ & 100 & 0.3 & 10.1 & 0.14 & $0.033\pm0.017$ & 0.092 \\
NGC\,1261  & $270.54$ & $-52.12$ & 11  & $23\times23$ & 100 & 0.3 &  8.4 & 0.01 & $0.024\pm0.008$ & 0.049 \\
NGC\,5927  & $326.60$ & $4.86$   & 3.5 & $23\times23$ & 100 & 0.3 &  8.4 & 0.45 & $0.059\pm0.031$ & 0.169 \\
NGC\,6656  & $9.89$   & $-7.55$  & 0.5 & $23\times23$ & 100 & 0.3 &  8.3 & 0.34 & $0.047\pm0.018$ & 0.100 \\
NGC\,6304  & $355.83$ & $5.38$   & 2.1 & $20\times20$ & 100 & 0.2 &  7.3 & 0.54 & $0.047\pm0.020$ & 0.109 \\
NGC\,6779  & $62.66$  & $8.34$   & 4.3 & $23\times23$ &  80 & 0.3 &  6.9 & 0.26 & $0.052\pm0.018$ & 0.093 \\
NGC\,7099  & $27.18$  & $-46.84$ & 3.9 & $20\times20$ &  80 & 0.3 &  6.0 & 0.03 & $0.030\pm0.010$ & 0.064 \\
NGC\,6934  & $52.10$  & $-18.89$ & 11  & $20\times20$ &  80 & 0.3 &  5.9 & 0.10 & $0.031\pm0.012$ & 0.069 \\
NGC\,4833  & $303.60$ & $-8.02$  & 1.4 & $20\times20$ &  80 & 0.3 &  5.6 & 0.32 & $0.042\pm0.015$ & 0.093 \\
NGC\,7006  & $63.77$  & $-19.41$ & 46  & $15\times15$ & 100 & 0.3 &  5.5 & 0.05 & $0.030\pm0.010$ & 0.055 \\
NGC\,6101  & $317.74$ & $-15.82$ & 7.3 & $20\times20$ &  80 & 0.3 &  5.4 & 0.05 & $0.026\pm0.009$ & 0.054 \\
NGC\,6723  & $0.07$   & $-17.30$ & 2.8 & $18\times18$ &  80 & 0.3 &  5.4 & 0.05 & $0.027\pm0.009$ & 0.051 \\
NGC\,6254  & $15.14$  & $23.08$  & 1.1 & $18\times18$ &  80 & 0.3 &  5.2 & 0.28 & $0.060\pm0.023$ & 0.117 \\
NGC\,6637  & $1.72$   & $-10.27$ & 5.2 & $18\times18$ &  80 & 0.3 &  5.1 & 0.18 & $0.031\pm0.007$ & 0.058 \\
NGC\,4590  & $299.63$ & $36.05$  & 3.4 & $18\times18$ &  80 & 0.3 &  5.0 & 0.05 & $0.046\pm0.017$ & 0.089 \\
NGC\,6584  & $342.14$ & $-16.41$ & 9.0 & $18\times18$ &  80 & 0.3 &  4.9 & 0.10 & $0.029\pm0.009$ & 0.055 \\
IC\,4499   & $307.35$ & $-20.47$ & 5.5 & $18\times18$ & 100 & 0.3 &  4.7 & 0.23 & $0.031\pm0.012$ & 0.064 \\
NGC\,6752  & $336.49$ & $-25.63$ & 1.0 & $18\times18$ &  80 & 0.3 &  4.4 & 0.04 & $0.030\pm0.012$ & 0.064 \\
NGC\,6426  & $28.09$  & $16.23$  & 11  & $15\times15$ & 100 & 0.3 &  4.3 & 0.36 & $0.047\pm0.022$ & 0.097 \\
NGC\,6624  & $2.79$   & $-7.91$  & 4.8 & $18\times18$ &  80 & 0.3 &  4.3 & 0.28 & $0.047\pm0.019$ & 0.114 \\
Palomar\,2 & $170.53$ & $-9.07$  & 25  & $15\times15$ &  80 & 0.3 &  4.2 & 1.24 & $0.161\pm0.090$ & 0.375 \\
\hline
\end{tabular}
\begin{list}{Table Notes.}
\item Col.~1: GC identification; Cols.~2 and 3: Galactic coordinates; Col.~4: Fraction of the 
half-maximum radius covered by WFC/ACS; Col.~5: RA and DEC grid; Col.~6: Minimum number of stars for 
a cell to be used in DR computation; Col.~7: Maximum allowed colour uncertainty; Col.~8: Number 
of stars remaining after applying the colour-magnitude filter and with $\sigma_{col}\le\sigma_{max}$; 
Col.~9: Foreground reddening (H10); Col.~10: Mean differential colour excess (values lower than
0.04 may be related to zero-point variations); Col.~11: 
Maximum value reached by the differential colour excess.
\end{list}
\end{table*}

\begin{table*}
\contcaption{}
\renewcommand{\tabcolsep}{2.1mm}
\begin{tabular}{@{}lrrccrccccc}
\hline\hline
GC & $\ell$ & $b$ & $\ss$ & $n_{RA}\times n_{DE}$ & $N_{min}$& $\sigma_{max}$ & $N_{CMD}$ & \EBV & \mEBV & \xEBV \\
   &($\degr$)&($\degr$)&&(cells)&(stars)&(mag)&($10^4$ stars)&(mag) & (mag) & (mag)\\
(1) & (2) & (3) & (4) & (5) & (6) & (7) & (8) & (9) & (10) & (11)\\
\hline
NGC\,6809  & $8.79$   & $-23.27$ & 1.0 & $20\times20$ &  80 & 0.4 &  3.8 & 0.08 & $0.027\pm0.010$ & 0.050 \\
NGC\,6681  & $2.85$   & $-12.51$ & 6.3 & $15\times15$ &  80 & 0.4 &  3.7 & 0.07 & $0.024\pm0.007$ & 0.041 \\
NGC\,6981  & $35.16$  & $-32.68$ & 9.0 & $16\times16$ &  80 & 0.4 &  3.2 & 0.05 & $0.018\pm0.008$ & 0.038 \\
NGC\,6218  & $15.72$  & $26.31$  & 1.3 & $18\times18$ &  80 & 0.3 &  2.8 & 0.19 & $0.027\pm0.008$ & 0.056 \\
NGC\,3201  & $277.23$ & $8.64$   & 0.8 & $18\times18$ &  80 & 0.4 &  2.7 & 0.24 & $0.057\pm0.019$ & 0.118 \\
Lynga\,7   & $328.77$ & $-2.80$  & 3.3 & $15\times15$ &  80 & 0.4 &  2.6 & 0.73 & $0.073\pm0.037$ & 0.191 \\
NGC\,6362  & $325.55$ & $-17.57$ & 1.8 & $18\times18$ &  80 & 0.4 &  2.5 & 0.09 & $0.025\pm0.008$ & 0.046 \\
NGC\,5466  & $42.15$  & $73.59$  & 3.5 & $18\times18$ &  80 & 0.4 &  2.4 & 0.01 & $0.024\pm0.009$ & 0.048 \\
NGC\,288   & $152.30$ & $-89.38$ & 2.0 & $15\times15$ &  60 & 0.4 &  2.1 & 0.03 & $0.047\pm0.018$ & 0.091 \\
Terzan\,8  & $5.76$   & $-24.56$ & 13  & $15\times15$ &  80 & 0.4 &  2.0 & 0.12 & $0.022\pm0.010$ & 0.055 \\
NGC\,6144  & $351.93$ & $15.70$  & 2.7 & $15\times15$ &  60 & 0.4 &  1.9 & 0.36 & $0.038\pm0.016$ & 0.079 \\
NGC\,2298  & $245.63$ & $-16.00$ & 5.5 & $15\times15$ &  60 & 0.4 &  1.8 & 0.14 & $0.062\pm0.024$ & 0.126 \\
NGC\,6652  & $1.53$   & $-11.38$ & 10  & $15\times15$ &  60 & 0.4 &  1.8 & 0.09 & $0.025\pm0.008$ & 0.047 \\
NGC\,6171  & $3.37$   & $23.01$  & 1.8 & $13\times13$ &  50 & 0.4 &  1.8 & 0.33 & $0.061\pm0.025$ & 0.112 \\
NGC\,6352  & $341.42$ & $-7.17$  & 1.4 & $15\times15$ &  60 & 0.4 &  1.7 & 0.22 & $0.042\pm0.018$ & 0.092 \\
NGC\,5053  & $335.70$ & $78.95$  & 3.3 & $15\times15$ &  60 & 0.4 &  1.7 & 0.01 & $0.029\pm0.011$ & 0.058 \\
NGC\,4147  & $252.85$ & $77.19$  & 20  & $13\times13$ &  50 & 0.4 &  1.7 & 0.02 & $0.019\pm0.009$ & 0.045 \\
Rup\,106   & $300.88$ & $11.67$  & 10  & $15\times15$ &  70 & 0.5 &  1.7 & 0.20 & $0.026\pm0.010$ & 0.051 \\
NGC\,6397  & $338.17$ & $-11.96$ & 0.4 & $13\times13$ &  60 & 0.4 &  1.3 & 0.18 & $0.019\pm0.009$ & 0.051 \\
Arp\,2     & $8.55$   & $-20.79$ & 8.0 & $15\times15$ &  60 & 0.4 &  1.0 & 0.10 & $0.026\pm0.010$ & 0.059 \\
NGC\,6838  & $56.75$  & $-4.56$  & 1.2 & $11\times11$ &  50 & 0.4 &  0.9 & 0.25 & $0.035\pm0.015$ & 0.074 \\
Terzan\,7  & $3.39$   & $-20.07$ & 14  & $11\times11$ &  50 & 0.4 &  0.8 & 0.07 & $0.033\pm0.011$ & 0.055 \\
NGC\,6717  & $12.88$  & $-10.90$ & 5.2 & $11\times11$ &  50 & 0.4 &  0.8 & 0.22 & $0.025\pm0.010$ & 0.057 \\
Palomar\,15& $18.88$  & $24.30$  & 20  & $11\times11$ &  50 & 0.5 &  0.7 & 0.40 & $0.033\pm0.014$ & 0.088 \\
Pyxis      & $261.32$ & $7.00$   & --- & $11\times11$ &  50 & 0.5 &  0.7 & 0.21 & $0.038\pm0.012$ & 0.070 \\
NGC\,6366  & $18.41$  & $16.04$  & 0.6 & $13\times13$ &  50 & 0.4 &  0.6 & 0.71 & $0.055\pm0.018$ & 0.112 \\
NGC\,6535  & $27.18$  & $10.44$  & 4.0 & $10\times10$ &  50 & 0.4 &  0.3 & 0.34 & $0.020\pm0.011$ & 0.044 \\
\hline
\end{tabular}
\end{table*}

\section{Results}
\label{results}

\begin{figure}
\resizebox{\hsize}{!}{\includegraphics{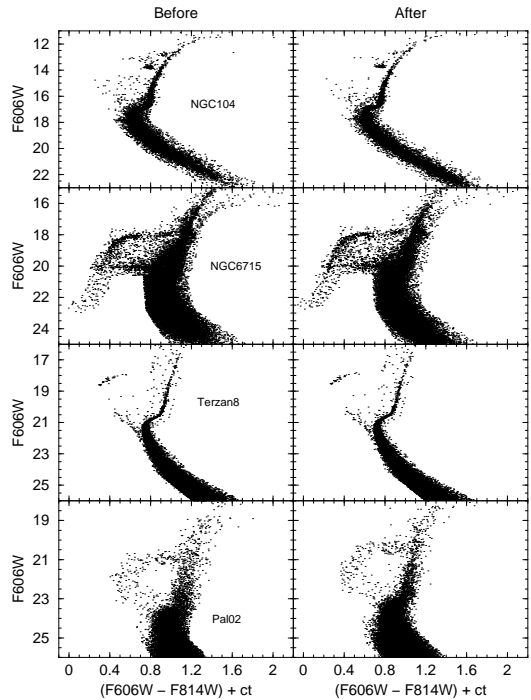}}
\caption{CMDs of selected GCs before (left panels) and after (right) 
differential-reddening correction. For visualization purposes, arbitrary shifts
have been applied to the colour axis.}
\label{fig5}
\end{figure}

The first step is to compute \eBV, the internal difference in \EBV\ of all cells with 
respect to the mean Hess diagram. Once the bluest cell is identified, we compute its 
difference in \eBV\ with respect to the other cells, thus giving rise to the actual 
distribution of DR values \dEBV\ in a given GC. Cells having a number of stars lower than 
$N_{min}$ (usually those lying at the image borders) are attributed a \dEBV\ corresponding 
to the average among the nearest neighbour cells. Finally, the colour and magnitude of all 
stars in a cell are corrected for the corresponding \dEBV, a process that is repeated for all 
cells in the grid. What emerges from this procedure is the corresponding DR map, such as those 
illustrated in Fig.~\ref{fig4}, and the DR-corrected CMDs (e.g. Fig.~\ref{fig5}).

Our approach computes colour shifts among stars extracted from different 
regions across the GCs. Thus, before proceeding into the analysis, we have to consider the potential 
effects of unaccounted-for zero-point variations across the respective fields. As discussed in 
\citet{Anders08}, stars extracted from wide-apart regions may present a difference in colour
(related to zero-point variation) of $\Delta(F_{606W} - F_{814W})\le0.04$, which our approach 
would take as a difference in reddening of $\dEBV\le0.04$. Correction for this residual
zero-point variation would require having access to a detailed map of the shifts for all GCs
in Table~\ref{tab1}. Since this information is not available, in what follows we either call
attention to or exclude from the analysis the GCs in which colour shifts due to zero-point 
variation may dominate over differential reddening. 

\begin{figure}
\resizebox{\hsize}{!}{\includegraphics{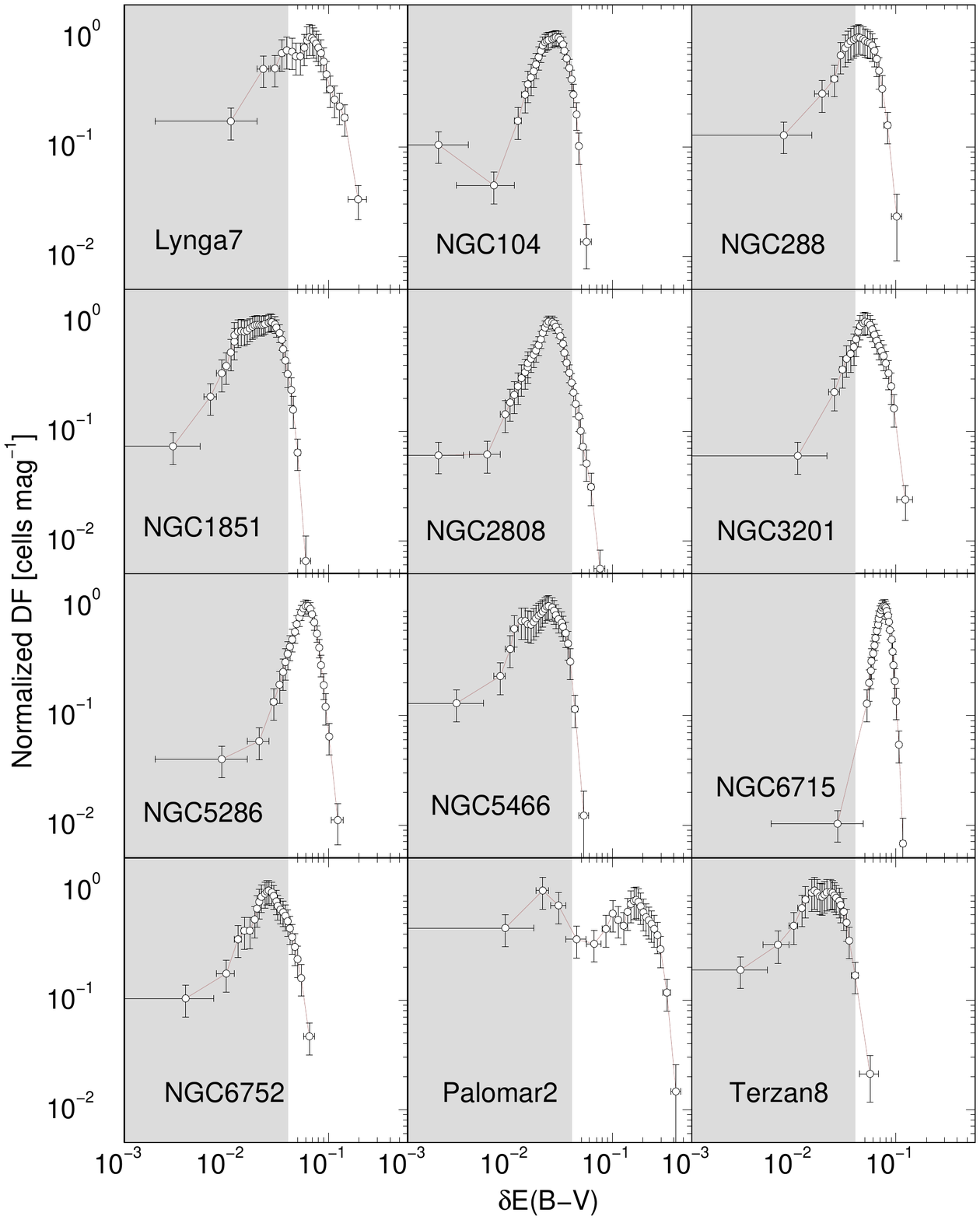}}
\caption{Distribution function of \dEBV\ for a representative GC sub-sample. For 
comparison purposes, the DFs are normalized to the peak. Values within the 
shaded area may be affected by residual zero-point variations. }
\label{fig6}
\end{figure}

The DR maps shown in Fig.~\ref{fig4} are typical among our GC sample. They show that extinction 
may be patchy (as in the centre of NGC\,6715) and scattered (as across Terzan\,8), rather uniform
(as in Palomar\,2, where it is apparently distributed as a thin dust cloud lying along the north-south 
direction), or somewhat defining a gradient (like in NGC\,104). However, most of the cells in 
NGC\,104 have $\dEBV\le0.04$ (Table~\ref{tab1} and Fig.~\ref{fig6}), which suggests that most
of the colour pattern across its field is extrinsic, probably related to the zero-point variation.

Turns out that our approach is capable of detecting some measurable amount of DR in the 66 GCs 
dealt with in this paper. To summarize this point, we show in Fig.~\ref{fig6} the distribution 
function (DF) of the \dEBV\ measured over all cells of selected GCs, from which the mean and
maximum DR values, \mEBV\ and \xEBV, can be measured. Some GCs, like NGC\,104, NGC\,2808, NGC\,5286, 
and NGC\,6715, have roughly log-normal DFs, while in others like Palomar\,2, the DFs are flat and 
broad. The majority ($67\%$) of the GCs have the mean DR characterized by $\mEBV\la0.04$ (a
range of values that may be mostly related to zero-point variations), with the highest value $\mEBV=0.161$ 
occurring in Palomar\,2. Regarding the maximum DR value, 67\% of the GCs have $\xEBV\la0.09$, with 
the highest $\xEBV=0.375$ again in Palomar\,2. For a more objective assessment of the range of \dEBV\ 
occurring in the GCs, we give in Table~\ref{tab1} the average and maximum values of \dEBV\ found for 
each GC.

By construction, when our approach corrects a CMD cell by the corresponding DR value, it 
shifts both the colour and magnitude of all stars in the cell towards smaller values of apparent
distance modulus and colour excess. Thus, the correction brings together the cell's photometric 
scatter that, after applying to all cells, makes the blue and red CMD ridges somewhat bluer. We 
note that this same effect shows up in the CMDs of Terzan\,5, for which a different approach was 
applied by \citet{Tz5} to correct for DR (see their Fig.~4). In addition, since the reddening 
vector for F606W and F814W is roughly directed along the MS, the colour and distance modulus 
shift should be more conspicuous in the evolved sequences, as shown in Fig.~\ref{fig5} for 
selected GCs.

\begin{figure}
\resizebox{\hsize}{!}{\includegraphics{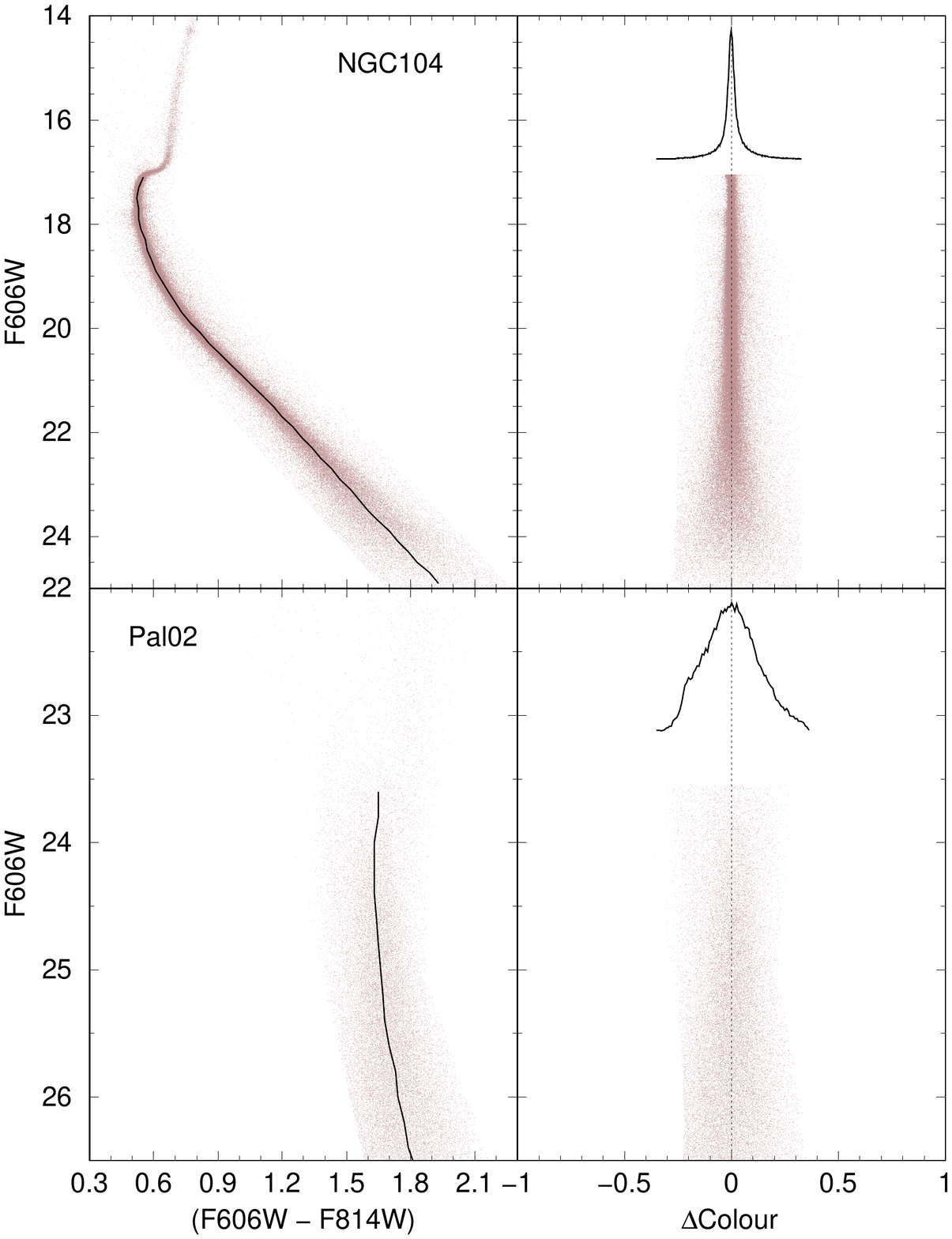}}
\caption{The fiducial line (solid curve, left panels) is used to rectify the main 
sequence (right panels), from which the residual colour distribution is computed
(inset in right panels).}
\label{fig7}
\end{figure}

\begin{figure}
\resizebox{\hsize}{!}{\includegraphics{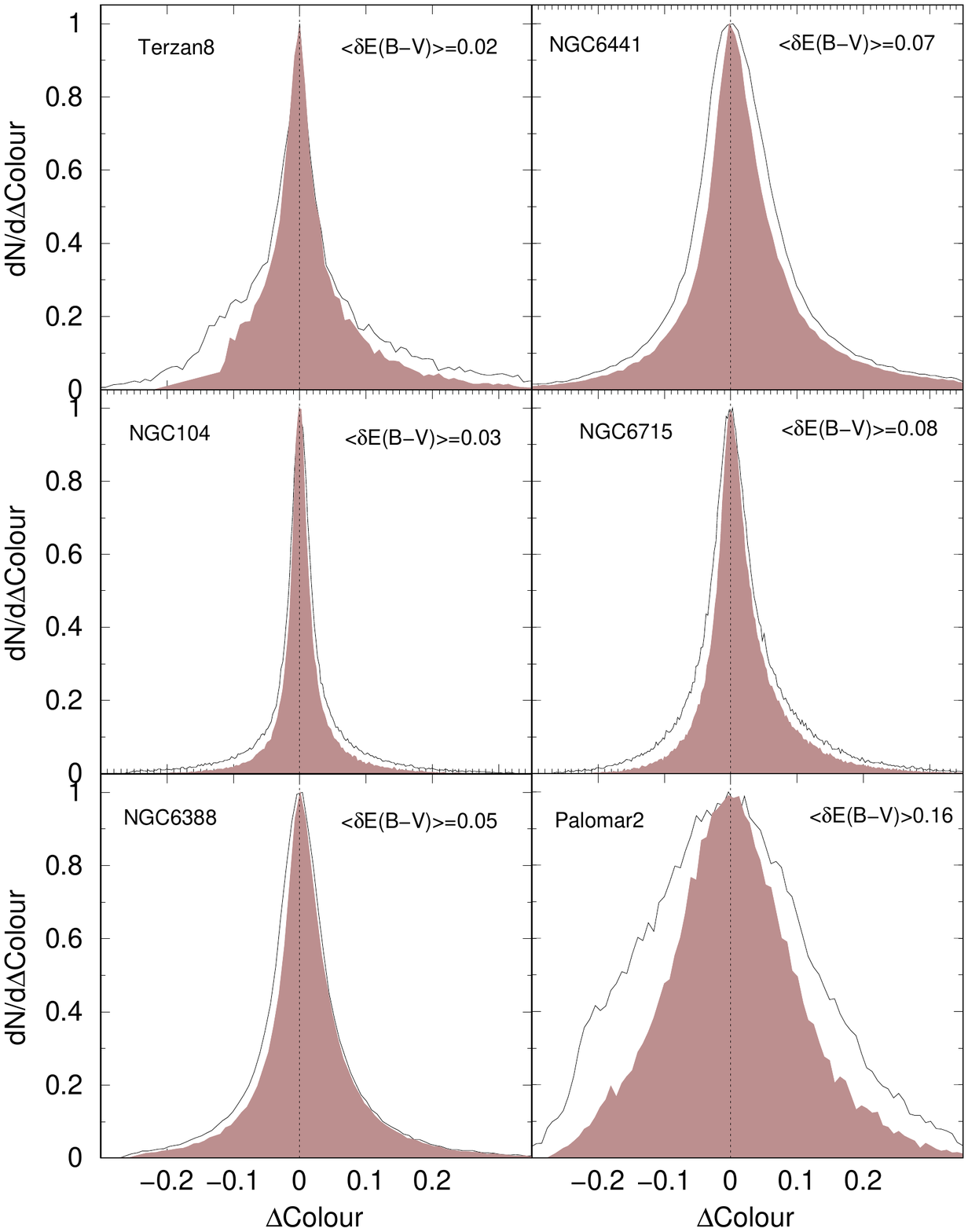}}
\caption{Residual colour distribution function of the main sequence of selected GCs before 
(solid curve) and after (shaded curve) differential-reddening correction.}
\label{fig8}
\end{figure}

For a more objective assessment of the changes introduced by our DR correction into the
CMDs, we examine the colour spread along the main sequence. The first step involves rectifying
the main sequence, which is achieved by subtracting - for each star - the colour corresponding
to the fiducial line, which basically traces the highest-density track of the CMD (or Hess diagram). 
In the present cases, we divide the main-sequence in bins 0.1 or 0.2 mag wide (depending on the
number of stars in each bin) and build the distribution function for the colour of all stars within
each bin. Then, we fit a Gaussian to the distribution function, from which we take the colour
corresponding to the highest stellar density - which represents the fiducial colour for the mid-point 
of the bin. After subtracting the fiducial line, we build the residual colour distribution function 
(i.e. the number-density of stars within a given colour bin) for the full magnitude range of the 
main sequence. This process is illustrated in Fig.~\ref{fig7} for NGC\,104 and Palomar\,2, previous 
to DR correction. The markedly different residual colour distributions (quite narrow for NGC\,104 
and broad for Palomar\,2) reflect the very different DR properties of both GCs (Table~\ref{tab1} 
and Fig.~\ref{fig6}). Figure~\ref{fig8} shows the residual colour distributions (restricted to the 
main sequence) before and after DR correction for selected GCs. The narrowing effect on the main 
sequence is conspicuous, especially for the highest differentially-reddened GC of our sample, 
Palomar\,2.     

\begin{figure}
\resizebox{\hsize}{!}{\includegraphics{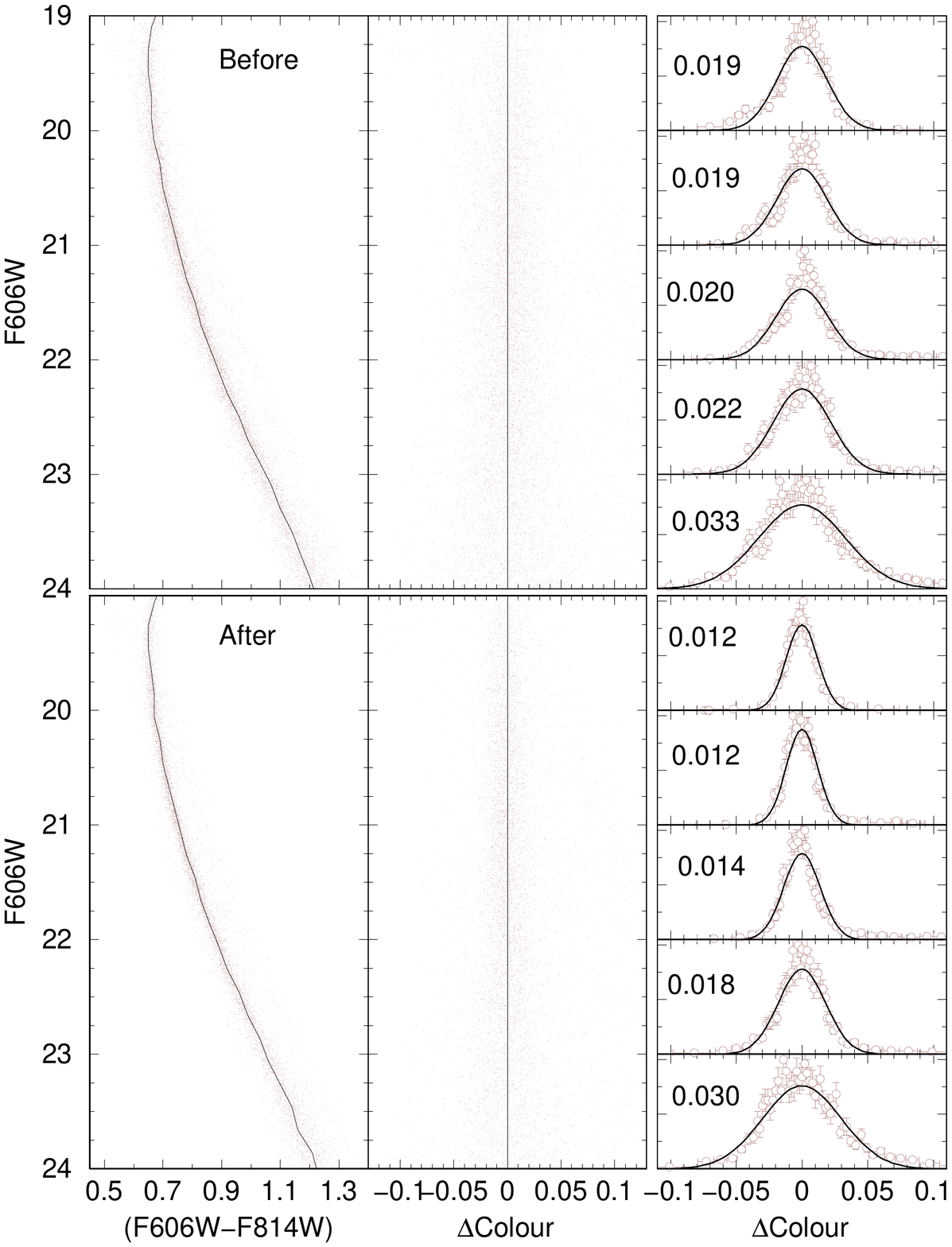}}
\caption{The major part of the main sequence of NGC\,2298 shown before (top panels) and after (bottom) 
DR correction. Subtraction of the fiducial line (left panels) allows visualization of the colour 
spread along the rectified main sequence (middle). The intrinsic colour dispersion is measured
in intervals of 1\,mag by means of Gaussian fit to the distribution function (right); the measured
dispersion ($\sigma_c$) is given in each panel (in all cases the uncertainty in $\sigma_c$ is 
$\le0.001$).}
\label{fig9}
\end{figure}

We further illustrate our approach in Fig.~\ref{fig9} with NGC\,2298 ($\EBV=0.14$, 
$\mEBV=0.062$, $\xEBV=0.126$), 
displaying the results in a similar fashion as in \citet{Milone2012a}, which also serves to
compare the different approaches more objectively. As expected, the DR correction decreases
the colour spread over most of the rectified main sequence. To better quantify the colour
spread, we build the colour distribution function for each interval $\Delta F606W=1$\,mag,
within the range $19 - 24$ (similar to that used by \citealt{Milone2012a}). Finally, we fit
a Gaussian to these functions, from which we measure the dispersion ($\sigma_c$) around the 
mean colour spread. Both before and after DR correction, $\sigma_c$ increases towards fainter
magnitudes; obviously, $\sigma_c$ values measured after DR correction are smaller than the 
respective ones before correction. Although following a similar pattern, our $\sigma_c$ values
are a little lower than those found by \citet{Milone2012a}. This difference is possibly
related to the fact the first step into the analysis involves excluding probable field stars 
(with colours excessively red or blue with respect to the expected GC evolutionary sequence) 
by using the colour-magnitude filter (Sect.~\ref{sample}). As an additional comparison, both 
the North-South gradient pattern indicated by our approach for the differential reddening in 
NGC\,6366 (Fig.~2 in \citealt{Campos13}) and the relative reddening values (Table~\ref{tab1})
are consistent with those found by \citet{Alonso97}, $\Delta\EBV\approx0.03$ between both
hemispheres.

An observational consequence of correcting a CMD for DR is a slight decrease in
the apparent distance modulus and colour excess by amounts related to \EBV\ according to the
adopted reddening vector. Consequently, the GC distance to the Sun should remain unchanged, 
since the distance decrease implied by the smaller apparent distance modulus is exactly 
compensated by the increase associated with the lower foreground colour-excess.

\begin{figure}
\resizebox{\hsize}{!}{\includegraphics{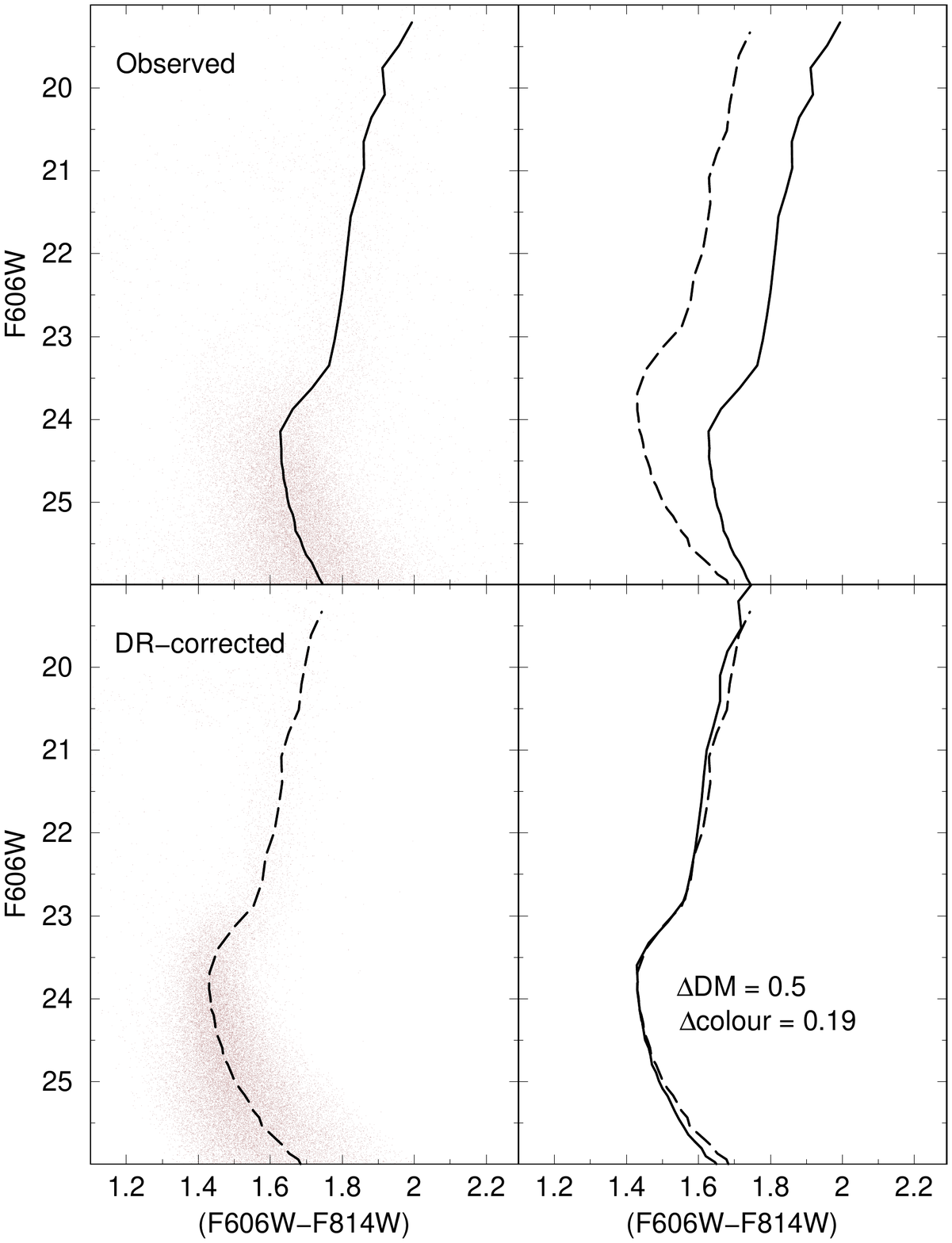}}
\caption{Differences in colour-excess and apparent distance modulus between the mean-ridge 
lines of the observed and DR-corrected CMDs of Palomar\,2 (left-panels) are significant
(top-right). They are minimized by decreasing the observed distance modulus by 0.5\,mag
and the foreground reddening by 0.19\,mag (bottom-right).}
\label{fig10}
\end{figure}

Figure~\ref{fig10} illustrates the above issue with Palomar\,2, the worst-case scenario in terms 
of mean, maximum (Table~\ref{tab1}) and distribution of DR values (Fig.~\ref{fig6}), as well as 
photometric scatter. We first build the mean-ridge line (MRL) for the MS, sub-giant and red-giant
branches both for the observed (top-left panel) and DR-corrected (bottom-left) CMDs. The differences
in apparent distance modulus and foreground colour excess become clear when both MRLs are placed
on the same panel (top-right). A near-match of both MRLs occurs when the apparent distance modulus 
and the colour excess of the observed CMD are shifted by $\Delta(m-M)_{F606W} = -0.5$ and 
$\Delta(F606W-F814W) = -0.19$, respectively. For these filters, the ratio between the DR-corrected
($d_c$) and observed ($d_o$) distances can be expressed as $\log(d_c/d_o) = \left(\Delta(m-M)_{F606W} - 
2.92\times\Delta(F606W-F814W)\right)/5$. For the MRLs of Palomar\,2, the shifts produce a small increase 
($\approx3\%$) in the distance, probably lower than the observational uncertainties. Thus, given the 
lower DR values and photometric scatter of the remaining GCs of our sample, the corresponding changes 
in distance should be lower than that of Palomar\,2.

Since the mean CMD morphology is essentially conserved by the DR correction, age and metallicity
determinations based on MRLs are expected to come up with unchanged results as well, but their
measured uncertainties should decrease, as the overall CMD scatter is reduced.

By definition, foreground reddening is the amount of extinction that affects uniformly all stars 
in a CMD. In this context, it is obvious that not all of the extinction needs to be actually 
located outside the cluster. For instance, suppose a GC affected by an internal DR distribution 
in which the minimum value is higher than zero and with no intervening extinction. Consequently, 
the minimum DR value would be mistaken as foreground. In this sense, our analysis above shows 
that part of the assumed foreground reddening may in fact be internal. In the case of Palomar\,2, 
the DR correction makes the overall CMD bluer by about $\Delta\EBV=-0.2$, thus implying a lower 
foreground reddening than that given in H10 (and reproduced in Table~\ref{tab1}). 

\begin{figure}
\resizebox{\hsize}{!}{\includegraphics{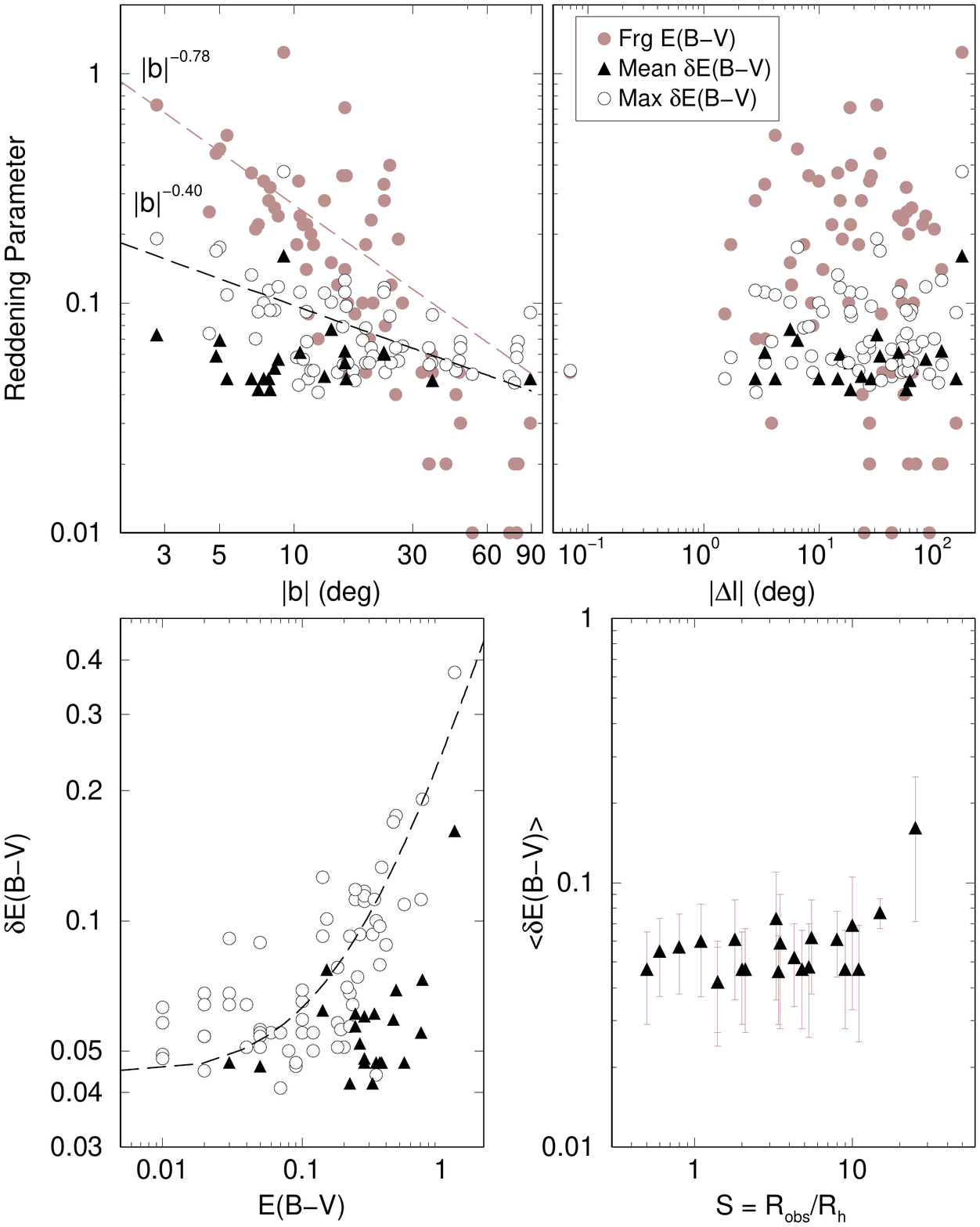}}
\caption{Dependence of foreground, mean and maximum differential-reddening values on the angular 
distance to the Galactic plane (top-left panel), and to the centre (top-right). The dependence 
on $|b|$ is characterized by power-law fits with the indicated slopes (dashed lines). \xEBV\ 
correlates with \EBV\ as a straight line (bottom-left). There is no correlation between \mEBV\ 
and the GC area covered by WFC/ACS (bottom-right). }
\label{fig11}
\end{figure}
  
Finally, in Fig.~\ref{fig11} we study dependences and correlations among the derived parameters. 
In this analysis we exclude the GCs having the maximum and mean DR lower than 0.04\,mag;
the final sub-samples contain 65 GCs with $\xEBV>0.04$ and only 22 with $\mEBV>0.04$. Both the 
foreground and maximum reddening parameters appear to anti-correlate with the distance to the 
Galactic plane ($|b|$) at different degrees (top-left panel); the mean-DR, on the other hand, 
presents an essentially flat distribution of values. As expected, the foreground absorption 
\EBV\ clearly increases towards the Galactic plane according to $\EBV=(1.59\pm0.44)\times|b|^{-(0.78\pm0.13)}$, 
with the mild correlation coefficient $CC=0.63$. A similar, but shallower dependence occurs for 
the maximum DR, $\xEBV=(0.24\pm0.05)\times|b|^{-(0.40\pm0.08)}$, with $CC=0.48$. 

%The nearly
%uniform behaviour of \mEBV\ with $|b|$, in particular, suggests that part of the measured DR 
%originates inside the clusters, otherwise, the mean DR should increase towards the plane with 
%a similar slope as the foreground \EBV. 

Both \xEBV\ and \mEBV\ are similarly uncorrelated with the angular distance to the Galactic 
centre ($|\Delta\ell|$); the same occurs with the foreground \EBV, but with the scatter having 
larger amplitude (top-right panel). Regarding the dependence on the foreground \EBV, the 
maximum DR values appear to increase with \EBV\ especially for $\EBV\ga0.2$ (bottom-left panel). 
Indeed, the best representation for \xEBV\ is a straight line, $\xEBV=(0.043\pm0.004) +
(0.20\pm0.01)\times\EBV$, with the strong correlation coefficient $CC=0.85$. The restriction 
to $\mEBV\ge0.04$ precludes a similar fit to \mEBV, but it also appears to increase for $\EBV\ga0.2$. 
This suggests that interstellar (external) reddening is the dominant source of DR.

%A possible interpretation is that reddening internal to the clusters is the main 
%source of DR when the foreground extinction is $\EBV\la0.2$.

The issue concerning the internal/external to the clusters nature of the differential-reddening 
source can be investigated from another perspective. Although the fraction of the cluster area 
covered by WFC/ACS $\ss\equiv R_{obs}/R_h$ varies significantly among our sample (Table~\ref{tab1}), 
its effect on the mean DR values is essentially negligible (Fig.~\ref{fig11}, bottom-right). A 
reasonable assumption is that, if the source of DR is essentially internal to the clusters, with 
the dust distribution following the cluster stellar structure, differences in \dEBV\ - and thus 
the mean DR - should increase with \ss. In this context, the lack of a systematic variation 
of \mEBV\ with \ss\ indicates that most of the DR source is external to the clusters, which is 
consistent with the conclusion of the previous paragraph.

\section{Summary and conclusions}
\label{Conclu}

In this work we derive the differential-reddening maps of 66 Galactic globular clusters from
archival HST WFC/ACS data uniformly observed (and reduced) with the F606W and F814W filters. 
The angular resolution of the maps are in the range $\approx7\arcsec\times7\arcsec$ to 
$\approx20\arcsec\times20\arcsec$. According to current census (\citealt{K49}, and references
therein), our sample comprises about 40\% of the Milky Way's globular cluster population,
which implies that the results can be taken as statistically significant.

We start by dividing the WFC/ACS field of view across a given GC in a regular grid with a
cell resolution that varies with the number of available stars. Next, we select a sub sample
of stars containing probable members having low to moderate colour uncertainty. Then, the 
stellar-density Hess diagrams built from CMDs extracted in different cells of the grid are 
matched to the mean one (containing all the probable-member stars available in the WFC/ACS 
image) by shifting the apparent distance modulus and colour excess along the reddening vector 
by amounts related to the reddening value \eBV\ according to the absorption relations in 
\citet{CCM89}. This is equivalent to computing the reddening dispersion around the mean, 
since a differentially-reddened cluster should contain cells bluer and redder than the mean. 
Finally, we compute the difference in \eBV\ between all cells and the bluest one, thus yielding 
the cell to cell distribution of \dEBV, from which we compute the mean and maximum values 
occurring in a given GC, \mEBV\ and \xEBV, respectively. This process also gives rise to the 
DR maps and the DR-corrected CMDs. 

We find spatially-variable extinction in the 66 GCs studied in this paper, with mean values 
(with respect to the cell to cell distribution) ranging from $\mEBV\approx0.018$ (NGC\,6981) 
up to $\mEBV\approx0.16$ (Palomar\,2). As a caveat, we note that values of $\dEBV\la0.04$ may 
be related to uncorrected zero-point variations in the original photometry. By comparing 
dependences of the DR-related parameters with the distance to the Galactic plane, foreground 
reddening, and cluster area sampled by WFC/ACS, we find that the main source of differential 
reddening is interstellar (external to the GCs). 

Regarding the CMD analyses based on mean-ridge lines (and theoretical isochrones), the DR 
correction decreases both the apparent distance modulus and foreground reddening values, but 
does not change the observed distance to the Sun, age and metallicity. Besides, the measured 
uncertainties in these fundamental parameters are expected to decrease, since the overall CMD 
scatter is reduced. Anyone interested in having access to the code and/or DR maps built in 
this paper - or having the data of specific GCs analysed by us - should contact C. Bonatto.

\section*{Acknowledgements}
We thank an anonymous referee for important comments and suggestions.
Partial financial support for this research comes from CNPq and PRONEX-FAPERGS/CNPq (Brazil). 
We thank A. Pieres for providing the routine to compute the mean-ridge lines. Data used in 
this paper were directly obtained from {\em The ACS Globular Cluster Survey: 
http://www.astro.ufl.edu/$\sim$ata/public$\_$hstgc/ }.

%--------------------------------- References -------------

\end{document}